\documentclass{epl}
\usepackage{amsmath}
\usepackage{graphics}

\newcommand{\beqa}{\begin{eqnarray}\label}
\newcommand{\beq}{\begin{equation}\label}
\newcommand{\eeqa}{\end{eqnarray}}
\newcommand{\eeq}{\end{equation}}

\title{Dynamics of Ising models coupled
microscopically to  bath systems }
\author{T. Plefka}
\institute{
 Department of Solid State Physics, TU Darmstadt, D
64289 Darmstadt, Germany} 
\begin{document}
\maketitle
\begin{abstract}
Based on the Robertson theory  the  nonlinear dynamics of general
Ising systems coupled microscopically to  bath systems is
investigated leading  to two complimentary approaches. Within the
master equation approach microscopically founded transition rates
are presented which essentially differ from the usual phenological
rates. The second approach leads to coupled equations of motion
for the local magnetizations  and the exchange energy. Simple
examples are discussed and the general results are applied to the
Sherrington-Kirkpatrick  spin glass model.
\end{abstract}
\section{Introduction} Ising spin models are probably the simplest interacting many
particle systems to which  many equilibrium  and non-equilibrium
phenomena found in nature can be mapped. In many cases the
analysis of these models exhibit the basic features and
consequently lead to a first understanding of the reality.
Therefore numerous investigations employing Ising models exist in
literature and are the subject of  research activities. In
addition to the questions in physics various problems  arising in
other fields (biology, social science, information theory, etc.)
are treated on the basis of Ising models.

Ising spin systems do not exhibit an intrinsic dynamics. Thus the
kinetics of these models is exclusively  caused  by couplings to
bath systems of the environment. Forty years   ago  Glauber
\cite{glauber} suggested   a phenomenological master equation  for
a one-dimensional chain of Ising spins with nearest neighbor
interactions. Generalized to other spin arrangements  and to other
types of interactions (for a review see \cite{kawa})  the approach
of Glauber has been nearly exclusively used to describe  the
dynamics of the Ising spin systems. Note that the above comment
also applies to the numerous Monte Carlo simulations performed on
various  models in the last decades \cite{binder}.

The transition rates of these phenomenological equations are
required to satisfy detailed balance which guaranties relaxation
to equilibrium  for non driven systems. But this requirement is
not sufficient to fix the rates completely.  Therefore ad hoc
assumptions have to be employed which lead to the claim
\cite{kawa} that  kinetic Ising models should not be regarded as
faithful representations of processes occurring in real systems.

It is  microscopic quantum-statistical investigations  which can
remove the  arbitrarily  of the phenomenological equations. The
tools for such a purpose are known since nearly half a century ago
\cite{first} and many general approaches exist. The application of
these general theories in Ising models, however, are rare. A first
approach of Heims \cite{heims}  uses a very special and thus
artificial spin-bath interaction. Further approaches, both based
on special realizations of the heat bath, are presented by Martin
\cite{martin} and recently by Park et al.\cite{park}.

All these microscopic approaches find transition rates different
from \cite{glauber}. A recent investigation for the linear
response of a general Ising model \cite{IIII} based on the Mori
theory results in the same conclusion.  It is the aim of the
present letter to calculate the transition rates under general
aspects and to work out the generic behavior for the general
non-linear case. Thus neither specific Ising models nor specific
realizations of the baths are presumed. The Ising spins may even
interact with more than one bath to include  problems like those
treated in \cite{schsch}.  Calculations are performed on the bases
of the Robertson theory \cite{rob,fs}, although equivalent result
\cite{just} are obtained by employing \cite{first} or the
Nakajima-Zwanzig  approach \cite{fs}.

With rare exceptions like the original problem  of Glauber
\cite{glauber}, analytical  solutions of the master equations  are
not known and therefore approximations have to be employed. The
typical procedure involves the elimination of fast degrees of
freedom in the master equation. This leads to a reduced
description and equations of motion result for the slow or
relevant variables. The latter equations, in many cases, can
directly be found by the Robertson theory. Thus both, the master
equation  and the equation of motion of such reduced descriptions
can  simultaneously be derived which justifies the above choice of
the method.

\section{The microscopic system:} A system of  $N$ quantum spins $
{\bf s} _i$ with $s={1\over2}$ is considered in the presence of
time dependent external fields $ H_i=H_i(t)$. The spins interact
via an arbitrary Ising spin-spin interaction $ J_{ij}(=J_{ji})$
and are described by the spin Hamiltonian
\begin{equation}\label{1}
{\cal H}^S =   -2 \sum_i H_i s^z_i +{\cal H}^{ex}\quad ,\; {\cal
H}^{ex}=- 2 \sum_{i,j}J_{ij}s_i^z s_j^z
\end{equation}
where $J_{ii}=0$ is presumed. Both quantum spin $1\over 2 $
operators $ \mathbf{s}_i$ with $\hbar=1$ and  Ising spins $S_i(=2
s_i^z)$ are used simultaneously.

The assembly of spins interact via  spin bath interactions ${\bf
B}_{\alpha i} {\bf s}_i$ with a system consisting of $N_B$
adiabatically isolated baths $ {\cal H}^{B}_\alpha $ \footnote{
This interaction represents the most general \textit{one} spin
coupling to the bath. Note that the contribution $B_i^z s_i^z$ has
no effect and that the coupling of \cite{heims} or similar two
spin couplings  will lead to spatial correlations via ${\cal
H}^{SB}$. Such a behavior may be possible but is certainly not the
generic case.}. Thus the total Hamiltonian of the system is
described by
\begin{equation}\label{3}
{\cal H}={\cal H}^{0} +{\cal H}^{SB}\:,\quad {\cal H}^0={\cal
H}^{S}+{\cal H}^{B}\:, \quad {\cal H}^{B}=\sum_{\alpha}{\cal
H}^{B}_\alpha\: ,\quad{\cal H}^{S B}= \sum_{\alpha
i}\,\frac{1}{2}( B^+_{\alpha i} s^-_i\,+\,B^-_{\alpha i}
s^+_i)\,+\,B^z_{\alpha i} s^z_i\:.
\end{equation}
During the dynamic evolution of the spin system  thermal
equilibrium is presumed for each individual   bath system $ {\cal
H}^{B}_\alpha $  leading to the statistical operator
\begin{equation}\label{15}
R^B =\,{\exp(-\sum \beta_\alpha {\cal H}_\alpha^B )} /{{\rm
Tr}\;\exp(-\sum \beta_\alpha {\cal H}_\alpha^B)}\quad.
\end{equation}
 These requirements on the bath systems can be realized by
sufficient large bath heat capacities and by faster relaxation of
the bath system. Thus $\tau_B\ll \tau_S $  where $ \tau_B$ and $
\tau_S $  are the characteristic relaxation times of the bath and
the spin system respectively. In general all the temperatures $
\beta_\alpha $ are different. Due to couplings to additional
reservoirs the  $ \beta_\alpha $ may even be time dependent. Thus
the $ \beta_\alpha (t) $ - together with $H_i(t)$ - are preset
quantities.  Both the $ \beta_\alpha (t) $ and the $H_i(t)$ are
presumed to have negligible variations on the fast time scale
$\tau_S $. \footnote{  Eq.(\ref{53}) is an approximation  which
requires this presumption   for the $H_i(t)$.}

With these requirements there is no need for an explicit
specification of the bath Hamiltonian ${\cal H}^B $ and the bath
operators $ {\bf B}_{\alpha i}$. As shown below, it is just the
absorptive part $\chi_{\alpha i}''(\omega)$ of the dynamic bath
susceptibility
\begin{eqnarray}\label{4}
  \chi_{\alpha i}(\omega) =\chi_{\alpha i}'(\omega) +i \chi_{\alpha i}''(\omega)
   = -i\int_0^\infty \langle [\,B^-_{\alpha i}, e^{i \mathrm{L}^B\,t} B^+_{\alpha i} \,
]\rangle_B\; e^{ i\omega t} \mathrm{d}t \quad,
\end{eqnarray}
which enters  the calculation for values  $ \omega\tau_S\approx1
$. Expectation values with respect to $R^{ B}$  are represented by
 $ \langle
\ldots\rangle_B$ and $\mathrm{L}^B$ denotes the Liouvillian
defined by $\mathrm{L}^B A= [ {\cal H}^ B,A] $. It is assumed that
$ \langle {\bf B}_{\alpha i} \rangle_{B}=0 $ holds, which can
always be achieved by a renormalization of the terms of the
Hamiltonian. Note that $ \omega\tau_S\approx1 $ implies $ \omega
\tau_B\ll 1 $ and the first term of the low frequency expansion of
$\chi_{\alpha i}''$ can be used. Apart from interesting exceptions
(compare \cite{park}) this yields in the generic case
\begin{equation}\label{7}
\chi_{\alpha i}''(\omega)\approx \textrm{const}\;
\omega\quad\textrm{for}\quad \omega \ll \tau_B^{-1}
\end{equation}
where the constant of proportionality  may  depend on
$\beta_{\alpha }$.
\section{Calculation:} Let  $ A^k( S_1,\ldots S_ N)$  be a fixed
but arbitrary set of variables which are functions   of the Ising
spins $ S_i$.  This set is assumed  to give a sufficient dynamical
description of some physical question and will be called
observation level according to \cite{fs}. Without further
specification of the observation level at this stage the  general
approach of Robertson \cite{rob,fs} is applied employing the usual
perturbation treatment up to the second order in $ {\cal H}^{SB} $
and the standard  Markovan approximation \cite{7}. As  all
variables $A^k$ commute with ${\cal H}^{0}$ this  leads to
\begin{equation}\label{50}
\frac{d}{dt}\langle A^k\rangle_t =-\int_0^ {\infty}
\,d\tau\langle[ \,U(t,t-\tau){\cal H}^{SB},[ {\cal H}^{SB},A^k]]
\rangle_t\quad,
 \quad U(t,t-\tau)= \exp (-i\tau[L^S(t)+L^B]) \eeq
 where the Liouvillians
 $   L^S  $ and    $ L^B $  are   associated to $  {\cal
 H}^S $ and to $  {\cal H}^B $ , respectively. The time dependent
expectation values $ \langle \ldots \rangle_t $ are performed with
the generalized canonical statistical operator
 \beq{54}
 R_t=R^S_t\: R^B
 \quad
 \mathrm{with}\quad
R^S_t={\exp(  V ) }/
 { {\rm Tr_S}\;\exp(V
 )}\quad  \textrm{and}\quad
 V=\sum_k \lambda^k_t A^k \quad
\end{equation}
where the time dependent Lagrange parameters $ \lambda^k_t$ are
implicitly determined by
 \begin{equation}\label{11}\langle A^k \rangle_t
={\rm Tr_S }\:A^k R_t^S \quad.
\end{equation}
Note that the system of eqs.(\ref{50}) and (\ref{11}) represent a
closed set of nonlinear differential eqs. for the expectation
values $ \langle A^k \rangle_t$ and the Lagrange parameters $
\lambda^k_t$. Thus together with the initial values for the $
\langle A^k \rangle_t$ (or with the initial values for $
\lambda^k_t$) the time dependence of all quantities  are
completely determined by this set of equations.

Analog to \cite{IIII}  eq.(\ref{50}) can be brought in a more
convenient form. Using again
 $\chi_{\alpha i}(\omega)=\chi_{\alpha i}^*(-\omega)$ and
$\chi_{\alpha i}''(\omega)= \pm \,\pi \,( e^{\pm \beta_\alpha
\omega}-1)\;\langle B_{\alpha i}^\mp\,\delta (
\mathrm{L}^B\pm\omega)\,B_{\alpha i}^\pm \rangle _{B}$ this leads
to
\begin{eqnarray}\label{53}
 \frac{d }{dt} \langle A^k\rangle_t& = \sum_{\alpha i}\big \langle
\Gamma_{\alpha i}\big\{ \tanh(\beta_\alpha H_i+\beta_\alpha X_i )
-S_i\big\}A^k_{[i]}\big\rangle_t \: \quad
\\\label{54}&\quad =
\sum_{\alpha i}\big \langle \Gamma_{\alpha i}\big\{ \tanh
(\beta_\alpha H_i+\beta_\alpha X_i )- \tanh V_{[i]}\big \}
A^k_{[i]}\big \rangle_t
\end{eqnarray}
with
\begin{equation}\label{55}
\Gamma_{\alpha i}  =\frac{ \chi_{\alpha i}''(2 H_i+2 X_i)}{2 \tanh
(\beta_\alpha H_i+\beta_\alpha X_i)}\quad .
\end{equation}
For all functions $ F $ of the $N$ Ising spins $ F_{[i]} $ is
defined as
\begin{equation}\label{60}
F_{[i]}=\frac{1}{2}{\rm Tr}_i \; S_i\; F ( S_1,\ldots,S_N)
\end{equation}
which enter in the calculation from  $ [s_i^\pm,F]=\mp 2 s_i^\pm
F_{[i]}$. The operators of the internal field at sites  $ i $ are
\begin{equation}\label{22a}
X_i=-{\cal H}^{ex}_{[i]}=\sum_j J_{ij}S_j\quad.
\end{equation}
In the  step leading to eq.(\ref{54}) the relation
\cite{ttcss,IIII}
\begin{equation}\label{22}
\langle \,S_i O \,\rangle_t= \langle \,O\,\tanh (V_{[i]})
\,\rangle_t
\end{equation}
 was used which holds for any
operator $O$  of the Ising spins not involving the spin $S_i$ .

 Eqs.(\ref{53}) and (\ref{54}) represent in a compact form the
most general results of this work. These results are valid for any
observation level which has to be specified for concrete physical
problems. As already pointed out, all slow variables of the
specific problem have  been included. One can also see that via
eq.(\ref{54}) a second requirement on the set of variables
results. Being, in principle, arbitrary  the initial value of
$R^S_{t=0}$  is assumed to have the special form of eq.(\ref{54}).
By an extension  of the observation  level, however, this second
requirement can always be satisfied for any given initial state.

Here two special observation levels will be discussed. First, the
observation level is analyzed which leads to the master equations
and  contains \textit{all} functions $F ( S_1,\ldots,S_N)$ of the
N Ising spins. Obviously all of the above requirements are
satisfied. The second observation level  is spanned by all Ising
spins $ S_1, \ldots,S_N$  and  by the interaction $ {\cal
H}^{ex}$. This choice  implies that all thermal equilibrium states
of the spin system are possible initial states. Consequently  this
observation level leads to a dynamical description of a standard
thermodynamic  system and is therefore  denoted as standard system
in the following.
\section{Master equation} Let the $|\bm{\sigma}\rangle$ be the
common eigenvectors of the $S_i$ where $
\bm{\sigma}=\{\sigma_1,\cdots \sigma_N\} $ denotes the
configuration of the spin system  with  the possible values
$\sigma_i=\pm 1 $. Then the observation level is spanned by all
the  $2^N$ projectors $|\bm{\sigma}\rangle\langle\bm{\sigma}|$.
Introducing  the notation $ \bm{\sigma}^{(i)}=\{\sigma_1,\cdots,-
\sigma_i,\cdots \sigma_N\} $ , the relation
$\big(|\bm{\sigma}\rangle\langle\bm{\sigma}|\big)_{[i]}=\sigma_i/2\,
\big(|\bm{\sigma}\rangle\langle\bm{\sigma}|\,+\,\,|\bm{\sigma}^{(i)}\rangle\langle\bm{\sigma}^{(i)}|
\big) $ holds. Eq.(\ref{53})  immediately leads  to the master
equation
\begin{equation}\label{m}
\frac{d \,}{d t}\,p^{\bm{\sigma}}\,= -\sum_i \Big( W_i^
{\bm{\sigma}}\,p^{\bm{\sigma}}-W_i^
{\bm{\sigma}^{(i)}}\,p^{\bm{\sigma}^{(i)}}\Big)
\end{equation}
for the occupation probabilities $ p^{\bm{\sigma}} =
\langle\bm{\sigma}|R_t^S|\bm{\sigma}\rangle$ .
 The transition rates $ W_{i}^{\bm{\sigma}}$ are given by
\begin{equation}\label{w}
 W_{ i}^{\bm{\sigma}} = \sum_\alpha\frac{C_{\alpha i}^{\bm{\sigma}}}{2}( 1-\sigma_i \tanh(\beta_\alpha
 H_i^{\bm{\sigma}})\big)=\sum_\alpha\frac{\chi_{\alpha i}''(2\,\sigma_i H_i^{\bm{\sigma}})}
 {4\big[\exp(2\beta_\alpha\,\sigma_i H_i^{\bm{\sigma}})-1\big]}
\end{equation}
with
\begin{equation}\label{g} C_{\alpha i}^{\bm{\sigma}}=\frac{
\chi_{\alpha i}''(2\, H_i^{\bm{\sigma}})}{4 \tanh (\beta_\alpha
H_i^{\bm{\sigma}} )}\quad \textrm{and}\:\quad H_i^{\bm{\sigma}}=
H_i+\sum_j J_{ij}\sigma_j\:.
\end{equation}
 \textit{The  rates $ W_i^{\bm{\sigma}}$} are consistent with
the results of \cite{park,IIII} but \textit{disagree with the
usual ad hoc assumption of  phenomenological kinetic Ising models}
\cite{glauber} which uses constant values of coefficient
$C_{\alpha i}^{\bm{\sigma}} $ independent of the configuration
$\bm{\sigma}$. Thus for a faithful representation of processes
occurring in real systems \textit{the full  $\bm{\sigma}$
dependence of the $C_{\alpha i}^{\bm{\sigma}} $ has  in  general
be used.}
\section{Standard system} This observation level will be
characterized by $V= -\lambda_{ex} {\cal H}^{ex}+\sum_i \lambda_i
S_i$ which leads to $V_{[i]}= \lambda_i +\lambda_{ex} X_i $.
Introducing the internal local-field probability functions
\begin{equation}\label{61}
w_i(x)=\langle
\delta(x-X_i)\rangle_t=\frac{\textrm{Tr}\,\delta(x-\sum_j
J_{ij}S_j) \exp (V)}{\textrm{Tr}\exp (V)}
\end{equation}
and employing again the relation (\ref{54}) the expectation values
of the  magnetizations $ m_i=\langle S_i \rangle_t$ and the
exchange energy $ U^{ex}=\langle {\cal H}^{ex} \rangle_t$
 can be written as (compare \cite{ttcss,IIII})
\begin{eqnarray}\label{65}
m_i\,=\int\textrm{d}x \, w_i(x) &\tanh (\lambda_i +
\lambda_{ex}x)\quad,\quad U^{ex}\,=
-\frac{1}{2}\sum_i\int\textrm{d}x \,x\, w_i(x) &\tanh (\lambda_i +
\lambda_{ex}x)\quad\quad
\end{eqnarray}
and the eqs.(\ref{53}) leads to
\begin{eqnarray}
&\frac{d \,}{d t}\,m_i\,= -\sum_\alpha\int\textrm{d}x \, w_i(x)
\frac{\chi''_{\alpha i}( 2 H_i + 2 x )}{2} \big [\frac{\tanh
(\lambda_i + \lambda_{ex}x) }{\tanh (\beta_\alpha H_i +
\beta_\alpha x)}
-1\big]\label{66}\\
&\frac{d \,}{d t}\,U^{ex} \,= \sum_{\alpha i}\int\textrm{d}x \,x\,
w_i(x) \frac{\chi''_{\alpha i}( 2 H_i + 2 x )}{2} \big
[\frac{\tanh (\lambda_i + \lambda_{ex}x) }{\tanh (\beta_\alpha H_i
+ \beta_\alpha x)} -1\big]\label{67}\:.
\end{eqnarray}
\textit{The set of nonlinear equations } (\ref{65}-\ref{67}) for
the dynamic variables $m_i$ and $U^{ex}$ together with the
associated time dependent Lagrange parameters $ \lambda_i$ and $
\lambda_{ex}$ \textit{is closed and is expected to give an
adequate description for many physical systems and situations.}

Note that the internal local-field distribution functions $ w_i(x)
$ govern the set of nonlinear equations of motion. Indeed all the
terms of these set of equations can explicitly be calculated from
the knowledge of the $ w_i(x) $. These findings generalize the
previous results, that the $ w_i(x) $  determine the statics
\cite{ttcss} and the dynamical linear response \cite{IIII} of an
arbitrary Ising model. In that previous work the distributions $
w_i(x) $ have explicitly been calculated for various specific
physical models including both ferromagnetic and disordered
systems. Consequently for all those models the nonlinear equation
of motion can easily be obtained from the eqs.(\ref{65}-\ref{67}).
\section{Application to the SK spin glass}  The latter procedure
will be illustrated for   infinite range spin glass model of
Sherrington and Kirkpatrick \cite{sk,mpv}. In the SK model  the
bonds $ J_{ij}$ are independent random variables with zero means
and standard deviations $N^{- {1\over 2}}$. This scaling fixes the
spin glass temperature to $T=1$. Employing  the modified TAP
approach\cite{II}
\begin{equation}\label{80}
w_i^{SK}(x)= \;\frac{1}{\sqrt{2 \pi \Delta}}\frac{ \cosh
(\lambda_i+ \lambda_{ex} x)} { \cosh (\lambda_i+
\lambda_{ex}H_i^{ex}) }\:\exp\big\{-\frac{(x-H_i^{ex})^2
+(\lambda_{ex} \Delta)^2}{2 \Delta}\big \}
\end{equation}
holds  \cite{IIII} with
\begin{equation}\label{81} H_i^{ex}=\sum_j
J_{ij}m_j -\lambda_{ex} m_i \Delta \quad \textrm{and\; with}\quad
\Delta =\frac{1}{N}\sum_i
\frac{(1-m_i^2)}{1+\Gamma^2\,\lambda_{ex}^{2}(1-m_i^2)^2} \quad,
\end{equation}
where $\Gamma $ is determined  from
\begin{eqnarray}\label{82}
\Gamma = 0\quad \quad \quad \quad\quad \quad \quad \quad \quad
\quad\quad \quad &\mathrm{f}\mathrm{or}&\quad
1-\frac{\lambda_{ex}^2}{N}\,\sum_i(1-m_i^2)^2\geq 0\\
1=\frac{1}{N}\sum_i
\frac{\lambda_{ex}^2(1-m_i^2)^2}{1+\Gamma^2\,\lambda_{ex}^{2}(1-m_i^2)^2}\qquad
 &\mathrm{f}\mathrm{or}&\quad
1-\frac{\lambda_{ex}^2}{N}\,\sum_i(1-m_i^2)^2\leq
0\label{83}\quad.
\end{eqnarray}
With the eqs.(\ref{80}-\ref{83}) the equations of motion
(\ref{65}-\ref{67}) are explicit provided the system of baths is
specified. For the case of one singular bath and identical spin
bath couplings  the sums over $\alpha$ drop out in
eqs.(\ref{66}-\ref{67})and only one bath temperature
 $\beta^B$ remains. The resulting  equations are expected to
 describe the complete non linear dynamics of the SK spin glass
 including  temperature quenches, memory and ageing effects \cite{young}.

A complete  discussion of these interesting effects is beyond the
scope of this work and is a subject of further research.  In the
following, I focus on the discussion  of the special situations
when the exchange system is in equilibrium with the bath, implying
$\lambda_{ex}=\beta_B$. This situation can  physically be realized
by allowing  only  changes of the  fields  $ H_i$. Setting
$\lambda_{ex}=\beta_B=\beta$ eq.(\ref{66}) yields Glauber type
equations of motion
\begin{equation}\label{90}
\dot{m}_i = - \gamma(\beta,H_i^{eff},\Delta) \, \big[m_i- \tanh
\beta H_i^{eff}\big ] \quad \quad ,\quad \quad H_i^{eff}=H_i
+\Sigma_j J_{ij}m_j-m_i\beta \Delta
\end{equation}
with the relaxation rates
\begin{equation}\label{91}
\gamma(\beta,H,\Delta)= \frac{\cosh \beta H }{\sqrt{2\pi}} \int
\textrm{d}z\,\exp\Big\{-\frac{\beta^2
\Delta+z^2}{2}\Big\}\;\frac{\chi_B''(2 H +{2 }\sqrt{\Delta }z) }{2
\;\sinh \beta (H + \sqrt{\Delta} z )}>0
\end{equation}
where $\Delta $ has to be determined from eq.(\ref{81}-\ref{83})
(with $\lambda_{ex}$ replaced by $\beta $). On a phenomenological
basis -  with $\gamma $ replaced by a constant -  equations of
motions of this kind have been used since the early days in spin
glass research \cite{dyn}. Moreover the author himself   had used
such equations to find numerically the solutions of the static TAP
equations \cite{II,III}. Note that eq.(\ref{91}) implies
relaxation to the static TAP equations independent of the values
of $\gamma(\beta,H_i^{eff},\Delta)$. Thus the  phenomenological
equations can be used instead of the microscopic results to
calculate static properties and these results justify   the
procedure used in \cite{II,III} to calculate the solutions of the
TAP equations.

For all dynamic  quantities, however, the full form of
$\gamma(\beta,H_i^{eff},\Delta)$ has to be used. First examples
are the dynamic susceptibility and the relaxation function as
presented already  in \cite{III}. Note that a result of
\cite{III}, the frequency-dependent shift of the cusp temperature
of the real part of the susceptibility, is basically a consequence
of $\gamma(\beta,H_i^{eff},\Delta)$. In this context the general
result \cite{fs} should be recalled  that the linear approximation
of the Robertson theory leads  the Mori theory. Consequently all
the linear response theory results of \cite{III} can also be
obtained  from the of eqs.(\ref{90}) and (\ref{91}) by linearizing
around the thermal equilibrium which can easily be checked.
\section{Discussion} The consequences
of the results of this letter are further illustrated on simple
physical situations. First one singular spin coupled to a bath
with temperature $\beta $ is considered. This implies no spin spin
interaction at all. Both the master equation and the standard
system approach of this letter leads to the equation of motion
$\dot{m} = - \gamma_1 ( m- \tanh \beta H )$ for the magnetization
$m$ with $ \gamma_1=\frac{1}{2}\chi_B''(2H)\coth \beta H $ . For a
static field the magnetization $m$ relaxes to the equilibrium
value $ \tanh \beta H $ with the well known (see e.g. \cite{fs})
longitudinal relaxation rate $ \gamma_1$. Note the latter result
is not found if instead of eq.(\ref{g}) the usual  ad hoc
assumptions of the kinetic Ising model are used.

Next the original system of Glauber, an infinite chain of Ising
spins with next neighbor ferromagnetic interactions is considered
which is coupled  to one bath with a temperature $\beta$. The
transition rates (\ref{g}) satisfy the detailed balance relation
and therefore relaxation to the thermal equilibrium results within
the master equation approach. The same consequences apply for the
standard system approach. Indeed the equilibrium values  of the
Lagrange parameter $ \lambda_i=H_i $ and $ \lambda_{ex}=\beta $
describe a fixed point of the eqs.(\ref{66}) and (\ref{67}). It is
obvious that nether the transition rates (\ref{g}) nor the details
of distribution functions $w_i$ enter in thermal expectation
values. Note that these arguments  apply to all undriven Ising
system which are coupled to one singular bath. Thus all this
systems relax to thermal equilibrium taking in account that the
fixed points are always stable. This can easily be checked by a
linear stability analysis.

For situations, however, where two or more bathes with different
temperatures $\beta_\alpha$ are present the values of stationary
solutions  depend on the transition rates or on the details of the
 distribution functions $ w_i(x) $. Thus for example the
non equilibrium  stationary  states of Schmittmann and Schm\"user
\cite{schsch} can not be accepted as a realistic solution for
their interesting model of a one dimensional Ising chain coupled
to two bath systems as the analysis is based on the
phenomenological  rates and not on the physical rates.

 Finally it is pointed out that the results of the present work will
 always effect the details of the dynamic evolution apart form situations
 where the hight temperature  approximation can be used.
In general  more complicated expressions are found compared to the
 phenomenological results. This even applies to the original
 model of Glauber.  For this kinetic Ising model  the equations of motion for the time
 correlation function partly separate which enables the
 explicit calculation of the complete dynamics. Such a simplifying separation does
 not occur for the microscopical master equation. Thus the solution  is
 not known and is hard to find as already pointed out in
 \cite{glauber}.
\section{Acknowledgments} Interesting discussions with B. Drossel,
 W. Just, G. Sauermann and F. Schm\"user are acknowledged.

\end{document}